\documentclass[]{aa}
\usepackage{graphicx}
\usepackage{txfonts}

\def\gsimeq
{\hbox{\raise0.5ex\hbox{$>\lower1.06ex\hbox{$\kern-1.07em{\sim}$}$}}}
\def\lsimeq
{\hbox{\raise0.5ex\hbox{$<\lower1.06ex\hbox{$\kern-1.07em{\sim}$}$}}}

\def\aj{{AJ}}

\def\apj{{ApJ}}
\def\apjs{{ApJS}}

\def\mnras{{MNRAS}}

\def\percm2{cm$^{-2}$}

\def\ltsima{$\; \buildrel < \over \sim \;$}
\def\simlt{\lower.5ex\hbox{\ltsima}}
\def\gtsima{$\; \buildrel > \over \sim \;$}
\def\simgt{\lower.5ex\hbox{\gtsima}}

\begin{document}
   \title{The X-ray and Radio Connection in Low-luminosity Active Nuclei}

   \subtitle{}

   \author{F. Panessa\inst{1}, X. Barcons\inst{1}, L. Bassani\inst{2}, M. Cappi\inst{2},
   	 F. J. Carrera\inst{1}, L.C. Ho\inst{3}, S. Pellegrini\inst{4}    
          }

   \offprints{Francesca Panessa\\ \email{panessa@ifca.unican.es}}

   \institute{Instituto de F{\'\i}sica de Cantabria (CSIC-UC),
   Avda. de los Castros, 39005 Santander, Spain
             \and INAF-IASF, Via P. Gobetti 101, 40129 Bologna, Italy
	     \and The Observatories of the Carnegie Institution of Washington, 813 Santa Barbara St.
	     Pasadena, CA 91101
	     \and Dipartimento di Astronomia, Universit\'{a} di Bologna, via Ranzani 1, 40127 Bologna, Italy  
}

   \date{Received / Accepted}
\authorrunning{F. Panessa et al.}


   \abstract{We present the results of the correlation between the nuclear 2-10 keV X-ray
   and radio (at 2~cm, 6~cm and 20~cm) luminosities for a well defined  sample of local
   Seyfert galaxies. We use a sample of low luminosity radio galaxies (LLRGs) for
   comparison. In both Seyfert and LLRGs samples, X-ray and radio luminosities 
   are significantly correlated over 8 orders of magnitude, indicating that 
   the X-ray and radio emission sources are strongly coupled. Moreover, 
   both samples show a similar regression slope, L$_{X}$
   $\propto$ L$_{R}^{0.97}$, but Seyfert galaxies are three orders of magnitude less
   luminous in the radio band than LLRGs. This suggests that
   either similar physical mechanisms are responsible for the observed emission or
   a combination of different mechanisms ends up producing a similar correlation slope. 
   Indeed, the common belief for LLRG is that both the X-ray and radio emission are likely
   dominated by a relativistic jet component, while in Seyfert galaxies the X-ray
   emission probably arises from a disk-corona system and the radio emission is
   attributed to a jet/outflow component.
   We investigate the radio loudness issue in the two samples and find that the Seyfert
   galaxies and the LLRGs show a different distribution of the radio loudness parameters.
   No correlation is found between the luminosity and the radio loudness, however
   the latter is related to the black hole mass and anti-correlated with the Eddington
   ratio. The dichotomy in the radio loudness between Seyfert and LLRG observed down to low Eddington ratios, 
   L$_{2-10 \mathrm{keV}}$/L$_{Edd}$ $\sim$ 10$^{-8}$, does not support the idea that the
   origin of the radio loudness is due to a switch in the accretion mode. 
    }

   \maketitle
%

\section{Introduction}

Recent developments have shown that all Active Galactic Nuclei (AGN) 
are radio sources at some level. Indeed, 
a high fraction of radio cores have been detected in 
radio-quiet AGN and their radio emission probably results from some sort of outflow
of radio plasma from the nucleus, in the form of jets, bubbles or
plasmoids (Wilson \& Ulvestad 1987, Pedlar, Dyson \& Unger 1985).
The issue that all AGN could host a jet and be synchrotron emitters,
is in apparent contrast with the sharp division observed between radio-loud 
and radio-quiet AGN. Radio-loud sources constitute 
only $\sim$ 10-20\% of the AGN population (Strittmatter et al. 1980, 
Kellermann et al. 1989). Optically selected AGN are historically divided in radio-loud
or radio-quiet depending on the value of the
radio loudness parameter R, defined as the ratio
between the monochromatic luminosities at radio and
optical frequencies (R $\equiv$ L$_{6 cm}$ / L$_{B}$).
Radio-quiet objects show values of R concentrated between 0.1-1,
while in radio-loud sources the R values range from 10 to 100 (Kellerman et al. 1989),
so that the boundary between the two classes is normally defined at R = 10
(Visnovsky et al. 1992, Kellerman et al. 1994).
Another criterion introduced by Miller, Peacock \& Mead (1990),
is based on the separation between the two classes
in the radio luminosity (P$_{6~cm}$ $\sim$ 10$^{25}$ W Hz$^{-1}$ sr$^{-1}$).
The existence of the bimodality in the AGN population  
has been recently questioned and 
ascribed to sample selection effects, i.e. in deep radio and optical 
surveys no bimodality is observed in the distribution of the
radio loudness parameter (White et al. 2000, Hewett et al. 2001, Cirasuolo et al. 2003).
However, on the base of a sample
of $\sim$ 10000 objects selected from SDSS and FIRST,
Ivezi{\'c} et al. (2004, 2002) still found a bimodal distribution of R.

Seyfert galaxies and low luminosity AGN 
(LLAGN\footnote{As LLAGN we refer to low luminosity Seyfert galaxies,
LINERs, and "transition nuclei" with spectra intermediate between
those of LINERs and HII regions.}) have been surveyed by VLA and VLBI observatories 
(Ho \& Ulvestad 2001, Nagar et al. 2002). 
Ho \& Ulvestad (2001) have shown that 
85\% of the Seyfert nuclei are detected at 6 cm. 
These authors found a wide range of radio powers 
and morphologies of the radio emission typical of that of 
a compact core (either unresolved or slightly resolved), 
occasionally accompanied by elongated, jet-like features.
Similarly, Nagar et al. (2002) have detected a high incidence
of pc-scale radio cores and sub-parsec jets in LLAGN.
However, the physics of jets in radio-quiet sources is still
unclear. Henri \& Petrucci (1997) and Malzac et al. (1998)
suggested that the initial part of a jet can produce relativistic
particles illuminating the disk, with the jet having
either bulk relativistic motion or containing very energetic particles.
Alternatively, the bulk velocities could be sub-relativistic
in a sort of 'aborted jet' which produce emission also at X-ray frequencies (Ghisellini et
al. 2004). This idea would be an alternative to the disk/hot corona model
which is commonly invoked to explain the observed X-ray emission in radio-quiet AGN
(Haardt \& Maraschi 1991).

The combination of data on both X-ray and radio frequencies
offers the opportunity to investigate the relationship between
the physical source of the X-ray emission, likely due to the disk-corona system
in Seyfert galaxies, and the radio source.
Monochromatic soft X-ray and radio luminosities have been correlated
in radio-quiet and radio-loud quasars (Salvato et al. 2004, Brinkmann et al. 2000, 
Canosa et al. 1999), confirming the idea
that the radio emission originates in a compact nuclear source
directly associated to the central engine.
Recently, Falcke et al. (2004) and Merloni et al. (2003) have combined
the X-ray, radio luminosity and black hole mass estimates and derived
the so-called fundamental plane of super-massive black hole activity. Falcke et al. (2004)
suggested that the radio through X-ray emission of LLAGN is attributed to synchrotron
emission from a relativistic jet, similarly to the scenario proposed for X-ray binaries (XRBs) 
in their low/hard state.
Moreover, the slope found for the X-ray and radio relation in their work
is consistent with that observed in stellar black hole systems
(Gallo et al. 2003). On the other hand, the fitting results by Merloni et al. (2003) 
require a radiatively inefficient accretion flow (Narayan \& Yi 1994,
Abramowicz 1997) as the origin for X-ray emission, and a relativistic jet
as the origin for the radio emission. Since the radiatively
inefficient accretion flow models apply for low Eddington ratios
(L/L$_{Edd}$ $\leq$ 10$^{-2}$), this theoretical view
can be suitable for LLAGN which are known to accrete at very low Eddington ratios.
However, several pieces of evidence have been accumulated showing that most 
of the low luminosity Seyfert galaxies and LLAGN share characteristics in common with their more luminous
counterparts, such as broad H$_{\alpha}$ emission lines and emission
line ratios characteristic of powerful AGN in the optical band (Ho, Filippenko, \& Sargent 1997a), 
and spectral slopes and FeK line features typical of luminous AGN in the X-ray band 
(e.g., Cappi et al. 2005, Ho et al. 2001); altogether pointing to similar emission processes and to radiatively efficient 
accretion mechanisms.

Indeed, an homogeneous X-ray study of a well defined sample of nearby 
Seyfert galaxies recently performed has confirmed a continuity between
the properties of low luminosity Seyfert galaxies and luminous AGN
(Cappi et al. 2006, Panessa et al. 2006). We have shown that
the distributions of the spectral parameters (such as the continuum slope,
the energy of the FeK line, the absorbing column density, etc.), 
in particular for type 1 objects,
are found to be within the range of values observed in luminous AGN.
In Panessa et al. (2006) we have also investigated the X-ray, optical
and black hole mass properties for a sample of Seyfert galaxies
with luminosities spanning from 10$^{37}$ erg s$^{-1}$ up to 10$^{43}$ erg s$^{-1}$.
Our study is consistent with the idea that Seyfert nuclei are 
a scaled-down version of luminous AGN. The strong correlations 
found between X-ray and optical emission
lines over nearly 8 orders of magnitude suggest
that the X-ray source and the accretion mechanism responsible for the
production of UV photons are strongly related, similarly to quasars and
independently of their nuclear activity level, and, therefore, 
consistently with radiative efficient accretion. 
In this paper, we have continued 
the investigation of the multi-wavelength properties
of the Palomar Seyfert sample, exploiting the availability of
high spatial resolution data both at X-ray and radio frequencies and correlating them
for the first time.

A description of the sample is given in Sec. 2; the X-ray versus radio 
at 20~cm, 6~cm and 2~cm correlation for Seyfert galaxies is presented in Sec. 3; 
the relation between X-ray and 6~cm radio luminosities 
is discussed in Sec. 4 where the Palomar Seyfert sample is also compared to a
sample of low luminosity radio galaxies; in Sec. 5 we investigated
the properties of the radio loudness parameter and in Sec. 6 we explored
its relationship with the black hole mass and the Eddington ratio. Finally, results are
discussed and summarized in Sec. 7.
All distance-dependent quantities
were transformed to our adopted cosmological parameters of 
H$_{o}$ = 75 km s$^{-1}$ Mpc$^{-1}$ and q$_{o}$ = 0.5.

\section{The sample}

The Seyfert sample studied here has been presented in 
a companion paper (Panessa et al. 2006). It comprises
47 out of 60 Seyfert galaxies from the Palomar 
optical spectroscopic survey of nearby galaxies 
(Ho, Filippenko, \& Sargent 1995) for which X-ray data are available. 
The sources are classified as type~2 (34 out of 60), type~1 (13 out of 60), 
and "mixed" Seyfert galaxies (8),
according to their position in the optical emission line diagnostic diagrams.
The "mixed" Seyferts are found near the boundary between
Seyfert and LINER, HII or transition classification, resulting in
a double classification (e.g., S2/T2, L2/S2, H/S2, etc.).
See Panessa et al. (2006) for a more detailed description of the sample.

\begin{figure*} 
\begin{center}
\parbox{18cm}{
\includegraphics[width=0.32\textwidth,height=0.25\textheight,angle=0]{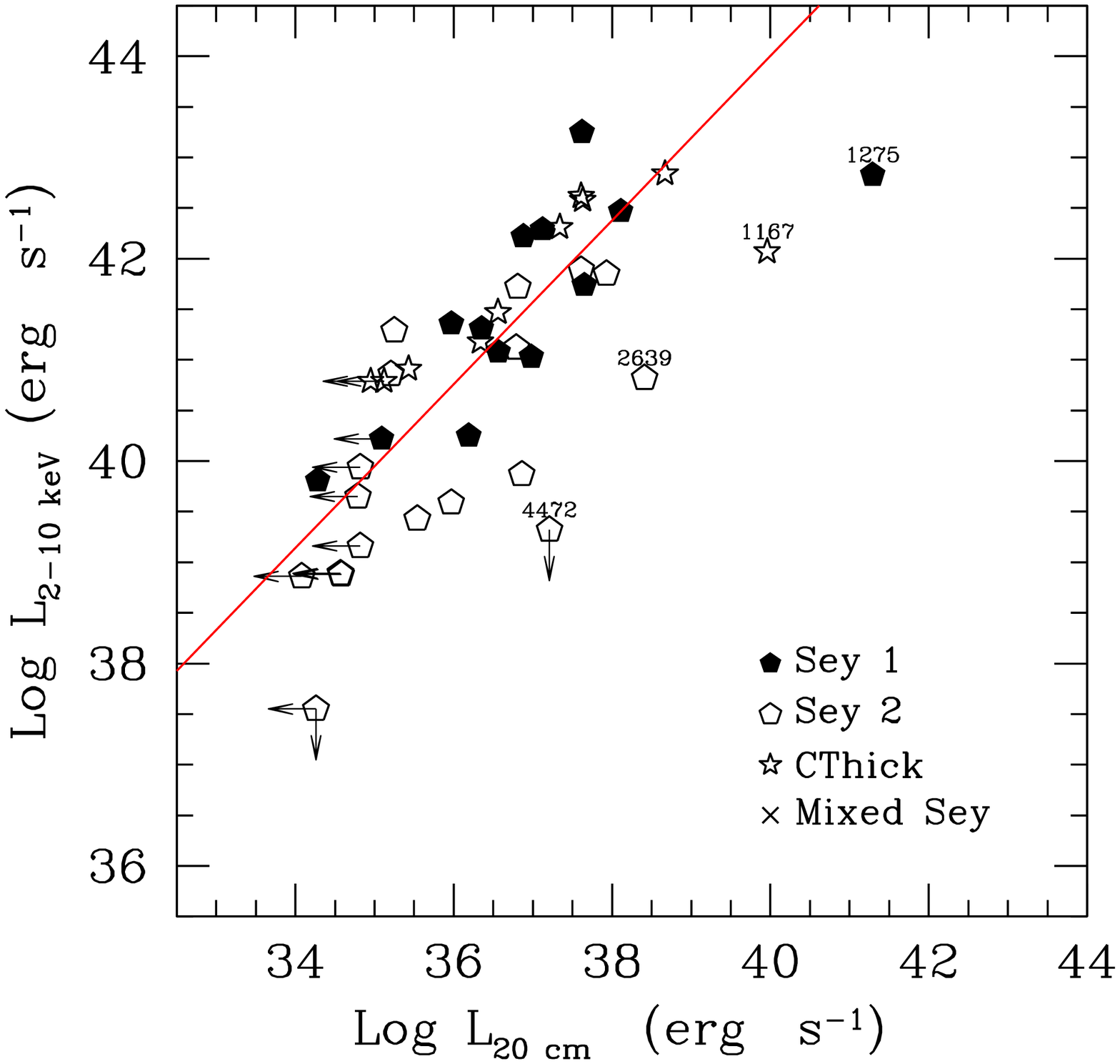}
\includegraphics[width=0.32\textwidth,height=0.25\textheight,angle=0]{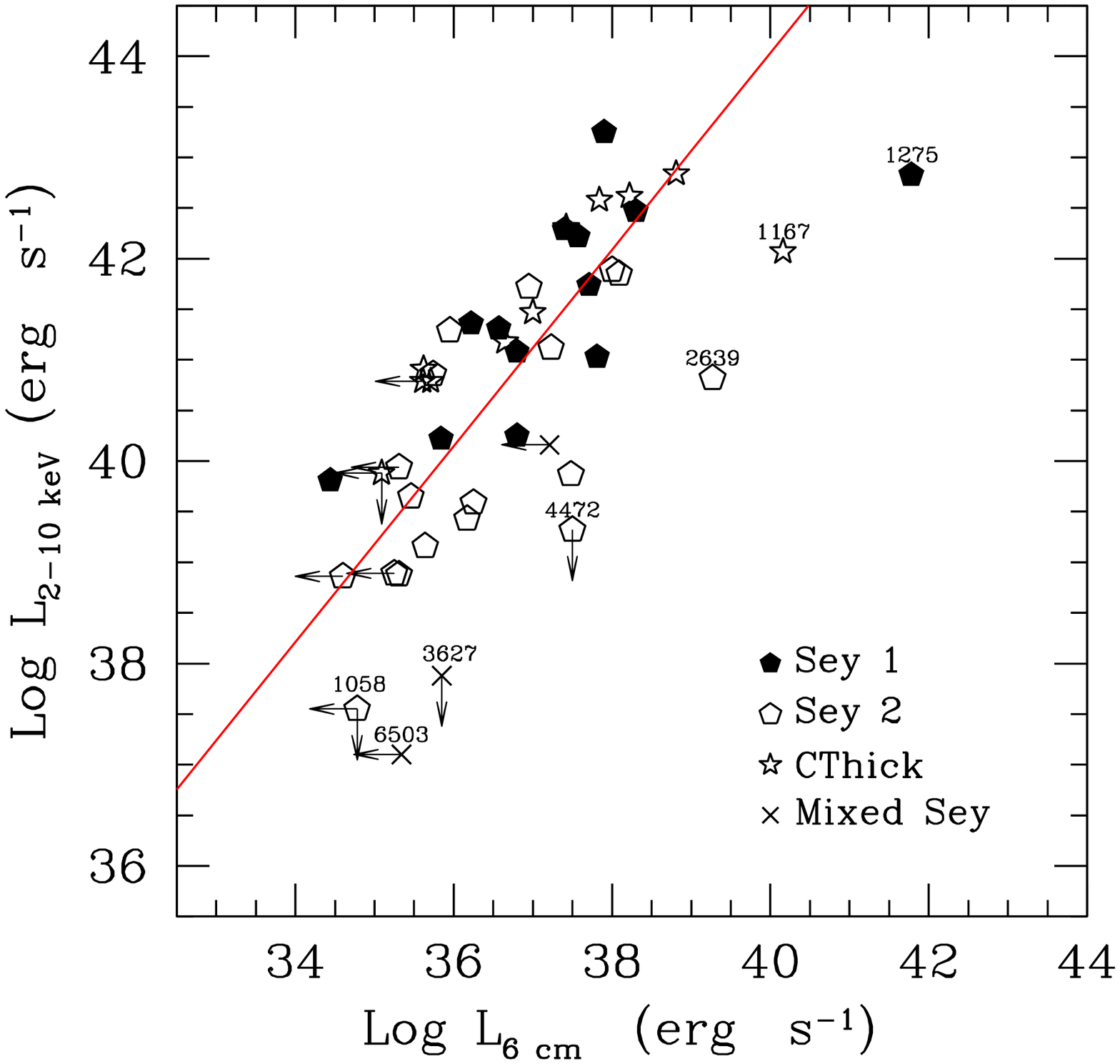}
\includegraphics[width=0.32\textwidth,height=0.25\textheight,angle=0]{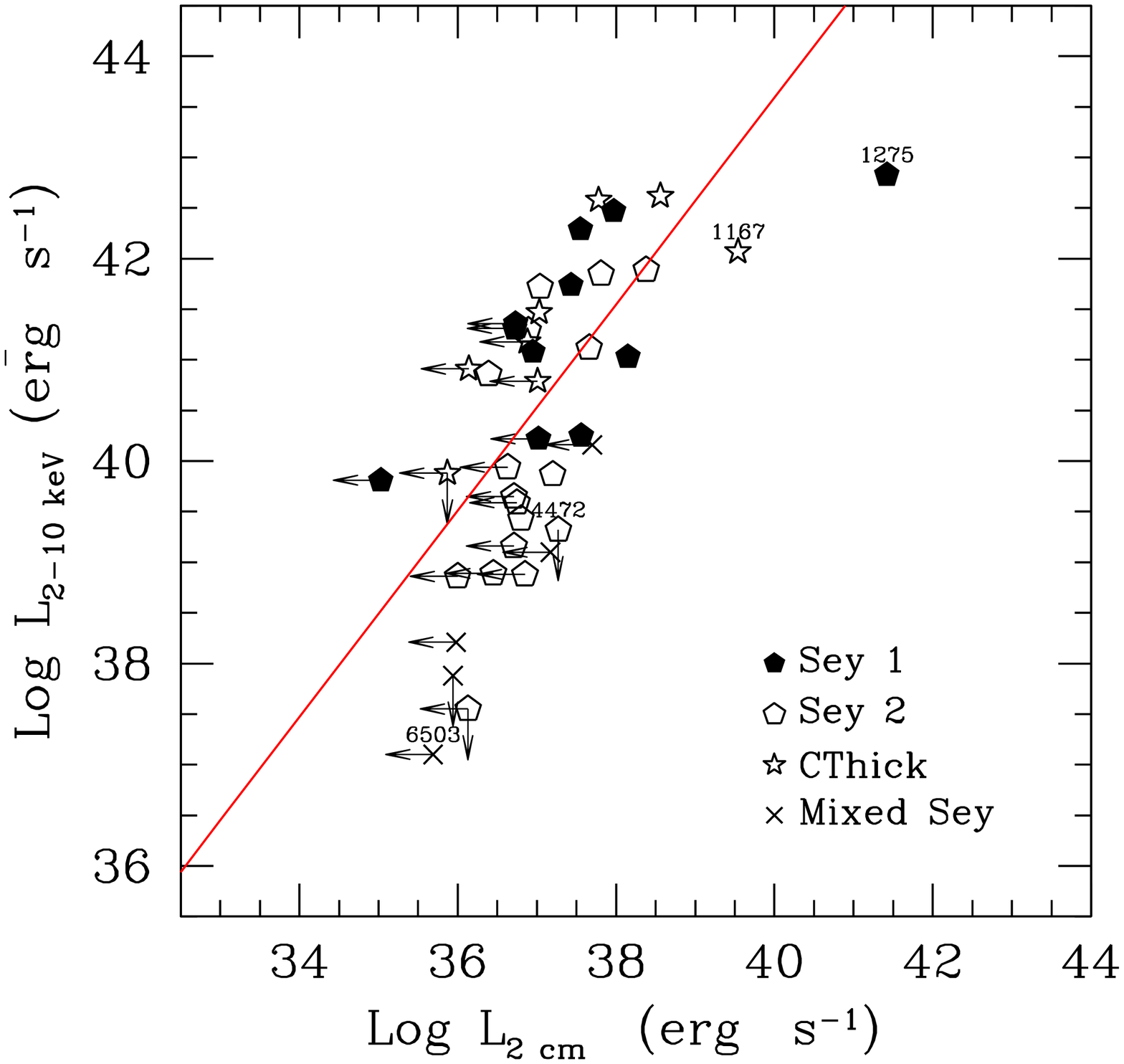}}
\caption{Intrinsic 2-10 keV luminosity
versus core radio luminosity at 20 cm (left), 6 cm (center) and 2 cm (right). Radio data
at 20 cm and 6 cm are taken from Ho \& Ulvestad (2001), at 2 cm are from Nagar
et al. (2005). Type 1 objects are plotted as filled polygons, 
type 2 as empty polygons, 'mixed Seyfert' objects
as crosses and Compton thick candidates as stars. The NGC~names
of a group of radio-loud Seyfert galaxies has been highlighted.}
\label{r1}
\end{center}
\end{figure*}   

\section{The L$_{X}$ versus L$_{R}$ correlation in the Palomar Seyfert sample}

The Palomar Seyfert sample has the merit of having
very accurate emission measurements in most of the spectral wavelengths.
In this section, we explore the relationship between the X-ray and 
radio luminosities, combining for the first time nuclear 
X-ray and core radio data obtained in recent surveys 
(Cappi et al. 2006, Panessa et al. 2006, 
Ho \& Ulvestad 2001, Nagar et al. 2002).

In Panessa et al. (2006), we have presented an X-ray study of this sample, using
the most recent {\it Chandra}, {\it XMM-Newton} and, only in a few cases, 
{\it ASCA} observations.
The nuclear 2-10 keV X-ray luminosities, obtained with minimal 
contamination by off-nuclear sources and diffuse emission,
have been corrected for Galactic and intrinsic absorption, the latter by using the 
measured X-ray column densities.
Among type 2 Seyfert galaxies, a sub-sample of 11 candidate Compton thick sources
($>$ 30\% of type 2 Seyferts) has been found. Since most of the 
intrinsic luminosity of Compton thick sources
is obscured from our line of sight and the intrinsic absorption 
is not measurable by data below 10 keV, we applied a correction factor
to the observed luminosities of these objects. For a detailed description
of the treatment of Compton thick candidate luminosities 
we refer to Cappi et al. (2006) and  Panessa et al. (2006).

Ho \& Ulvestad (2001) have undertaken a radio continuum 
survey of 52 Palomar Seyfert galaxies using the Very Large Array (VLA). 
The observations were made at 6 cm and at 20 cm with an 
angular resolution of $\sim$ 1$^{\prime\prime}$. 
Nearly 47\% of the Seyfert galaxies from the Palomar sample have
also been detected by a VLA survey at 2~cm, performed with a 
resolution of 0.15$^{\prime\prime}$ (Nagar et al. 2005).
Moreover, we have included data at 6~cm from Filho et al. (2002) for 
a group of four 'mixed' sources,
NGC~3489, NGC~3627, NGC~6482 and NGC~6503, not included in the  
Ho \& Ulvestad (2001) sample.

In Figure~\ref{r1}, we plot 
the absorption corrected 2-10 keV luminosity versus the core radio luminosity
at 20 cm (left figure), 6 cm (center) and 2 cm (right). 
In each case, we have performed a partial Kendall $\tau$ correlation
test, which computes partial correlation
coefficient and significance for censored data 
using three variables, where we took the distance as the third variable 
(Akritas \& Siebert 1996).
The X-ray versus radio luminosity correlation is
highly significant at all radio frequencies considered (with a probability
greater than 99.9\%). In Table~\ref{stat} we also report the Spearman's rho
correlation coefficient for comparison.

We characterize the above correlations by using the Schmitt method, 
which treats censored data in both variables
(Schmitt et al. 1985, the algorithm is implemented in the ASURV package, 
Isobe, Feigelson \& Nelson 1986). We report the Schmitt's slope
and intercept in Table~\ref{stat} as well. 

The X-ray versus radio luminosity correlations in Figure~\ref{r1} show
a group of outliers, i.e. sources that
present an excess in the radio emission with respect
to the average X-ray/radio ratio shown by the sample.
Interestingly, these sources have been classified in previous works
as radio-loud objects, as we briefly discuss in the following. 
The type 1 Seyfert having the highest radio 
luminosity in each plot is NGC~1275;
this is a radio-loud source (3C 84) having well known jets 
studied in detail (Ho \& Ulvestad 2001 and references therein).
NGC~1167 (a likely Compton thick candidate) 
is a well-known radio source (4C +34.09) and it has been
extensively studied at radio frequencies (Komossa et al.1999).
NGC~2639 is a radio-loud galaxy with an extended radio morphology. 
When observed at higher spatial resolution (VLBI), 
NGC~2639 is found to be one of the rare examples of a radio-emitting 
spiral galaxy with a VLBI core source (Hummel et al. 1982).
Finally, NGC~4472, an early-type galaxy, 
is also a radio-loud source (Caon et al. 1994).

Another small group of outliers is visible in the L$_{X}$ vs. L$_{6 cm}$ plot 
(central panel of Figure~\ref{r1}), at X-ray luminosities lower than 10$^{38}$ erg s$^{-1}$. 
Two out of these three sources are classified as 'mixed Seyfert',
i.e. NGC~3627 and NGC~6503,
while NGC~1058, which is a type 2 Seyfert galaxy, has an upper limit to the X-ray luminosity.
As already discussed in Panessa et al. (2006), these sources
show a peculiar behaviour and their emission could have
a starburst origin rather than an AGN nature.

In Table~\ref{stat}, we have reported the best fit linear regression line
obtained at each radio frequency either by including the above radio-loud sources 
and by excluding them; the latter is the fit plotted in Figure~\ref{r1}.
Note that the slope of the regression line becomes steeper when considering
higher radio frequencies, from 20~cm to 6~cm and 2~cm. The observed correlations suggest 
that the X-ray and radio emission mechanisms are strongly related in Seyfert galaxies,
as further discussed in the next section.

\begin{table*}
\small{
\caption{\bf Correlation statistics in luminosities}
\label{stat}
\begin{center}
\begin{tabular}{lcccccccccc}
\hline
\hline
\multicolumn{1}{c}{Variables} &
\multicolumn{1}{c}{N} &
\multicolumn{1}{c}{X(ul)} &
\multicolumn{1}{c}{Y(ul)} &
\multicolumn{1}{c}{XY(ul)} &
\multicolumn{1}{c}{Spearman} &
\multicolumn{1}{c}{$\tau$} &
\multicolumn{1}{c}{$\sigma$} &
\multicolumn{1}{c}{Prob.} &
\multicolumn{1}{c}{a} &
\multicolumn{1}{c}{b} \\
\multicolumn{1}{c}{(1)} &
\multicolumn{1}{c}{(2)} &
\multicolumn{1}{c}{(3)} &
\multicolumn{1}{c}{(4)} &
\multicolumn{1}{c}{(5)} &
\multicolumn{1}{c}{(6)} &
\multicolumn{1}{c}{(7)} &
\multicolumn{1}{c}{(8)} &
\multicolumn{1}{c}{(9)} &
\multicolumn{1}{c}{(10)} &
\multicolumn{1}{c}{(11)} \\
\hline
\\
 Seyfert Galaxies \\
 \\
 \hline
log L$_{X}$ vs log L$_{20 cm}$  &     41  & 2  &   2 & 8     &  0.79	& 0.50  & 0.10  &7.15E-07 & 0.62$\pm$0.04  &  18.17$\pm$1.46 \\
 			       &37$^{a}$  & 8  &   1 & 2     &  0.87    &      &       &          & 0.81$\pm$0.02  &  11.60$\pm$0.72   \\
log L$_{X}$ vs log L$_{6 cm}$  &      45  & 6  &   2 & 1     &  0.78	& 0.43 & 0.09  &$<$1.0E-08& 0.71$\pm$0.01  &  14.62$\pm$0.28 \\
                               &40$^{b}$  & 5  &   1 & 1     &  0.84    &      &       &          & 0.97$\pm$0.01  &  5.23$\pm$0.28	\\
log L$_{X}$ vs log L$_{2 cm}$  &      41  & 15 &   2 & 2     &  0.68	& 0.36 & 0.07  &7.15E-07  & 0.80$\pm$0.05  &  10.77$\pm$1.87 \\
                               &37$^{c}$  & 14 &   1 & 2     &  0.67    &      &       &          & 1.02$\pm$0.03  &  2.79$\pm$0.96\\
\hline
\\
Low-Luminosity radio galaxies \\
 \\
 \hline
log L$_{X}$ vs log L$_{6 cm}$  &      33  & 6  &   5 & 1     &  0.76	& 0.42 & 0.10 &1.67E-05  & 0.97$\pm$0.02  &  2.42$\pm$0.92 \\
\hline
\hline
\end{tabular}
\end{center}
Notes: Statistical properties of the 2-10 keV X-ray luminosity 
versus radio luminosities at 20 cm, 6 cm and 2 cm; Col. (1): Number of sources;
Col. (2)-(3)-(4): Number of upper limits in variable X, Y and both;
Col. (5) Spearman's rho correlation coefficient; Col. (6)-(7)-(8) Kendall's $\tau$ correlation coefficient,
the square root of the variance, $\sigma$, and the associated probability P for accepting the null hypothesis
that there is no correlation; Col. (9)-(10): Correlation coefficient of the best fit linear 
regression line calculated using Schmitt binned linear regression method, Y= a $\times$ X + b. 
Radio-loud sources excluded: $^{a}$ NGC~1275, NGC~1167, NGC~2639 and NGC~4472; $^{b}$ 
NGC~1275, NGC~1167, NGC~2639, NGC~4472 and NGC~6503; $^{c}$ NGC~1275, NGC~1167, NGC~4472 and NGC~6503.}
\end{table*}

\section{The L$_{X}$ versus L$_{6 cm}$ correlation}

In this section we focus on the L$_{2-10 keV}$ versus L$_{6 cm}$ correlation
and compare the Palomar Seyfert galaxy sample with a sample of low luminosity radio galaxies
taken from two different catalogues.

Balmaverde \& Capetti (2006) have selected a sample of 116 early type galaxies
(morphological Hubble type ${\it T}$ $\leq$ -1) detected
in VLA surveys limited in flux at $\sim$ 1 mJy. Using HST data for 65 out of 116 sources,
the sources have been separated in "core" and "power-law"
galaxies according to their nuclear brightness profiles.
Here, we focus our study on these "core" galaxies
which have been optically classified as type 1, type 2 Seyferts and LINERs, when an optical
classification is available.
The radio maps of the 29 "core" radio galaxies indicate collimated outflows,
which suggest that miniature radio-galaxies are probably residing
in these sources (Balmaverde \& Capetti 2006).
Nuclear Chandra X-ray luminosities are available for 20 out of 29 sources. 

In addition, we also considered a sample of 16 low luminosity radio galaxies 
from the 3C/FRI sample (Chiaberge, Capetti \& Macchetto 2005),
for which Chandra X-ray luminosities are reported in Balmaverde \& Capetti (2006).
The optical classifications of the sources belonging to this sample
span from Seyfert galaxies (both of type 1 and type 2) to LINERs.
Note that NGC~4261, NGC~4374 and NGC~4486 are both in the 
"core" radio galaxy sample and in the 3C/FRI sample, so when computing the best fit
regression lines we considered them only once. 
Also note that the type 1 Seyfert galaxy NGC~1275 (3C84) belongs also to 
the 3C/FRI sample, while the type 1.9 Seyfert NGC~4168 (UGC~7203) 
is also included in the "core" galaxy sample. 
Here we treat the core galaxy and the 3C/FRI galaxy samples 
as a unique sample of low-luminosity radio galaxies (LLRGs).

\begin{figure*} 
\begin{center}
\parbox{18cm}{
\includegraphics[width=0.50\textwidth,height=0.35\textheight,angle=0]{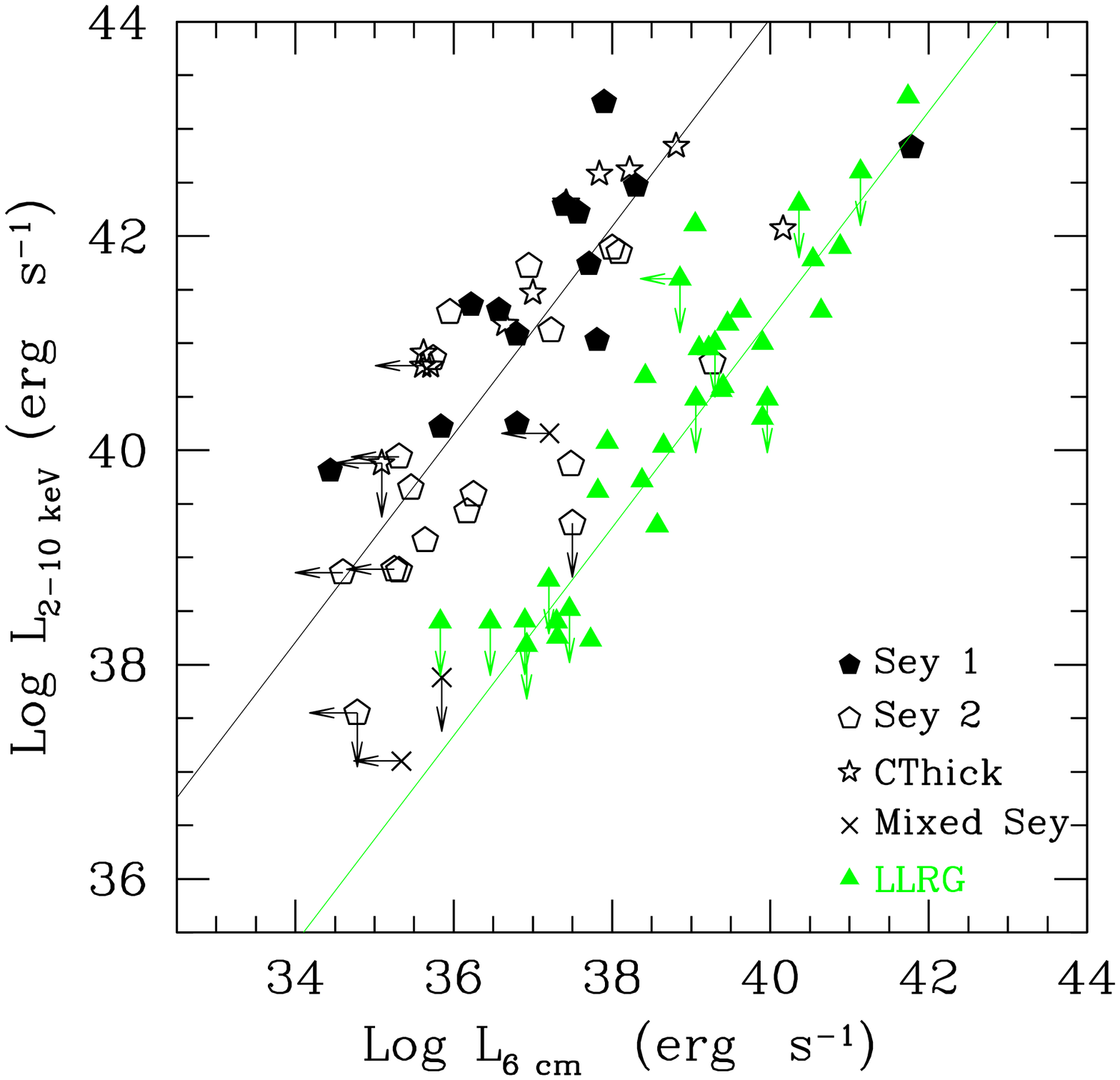}
\includegraphics[width=0.50\textwidth,height=0.35\textheight,angle=0]{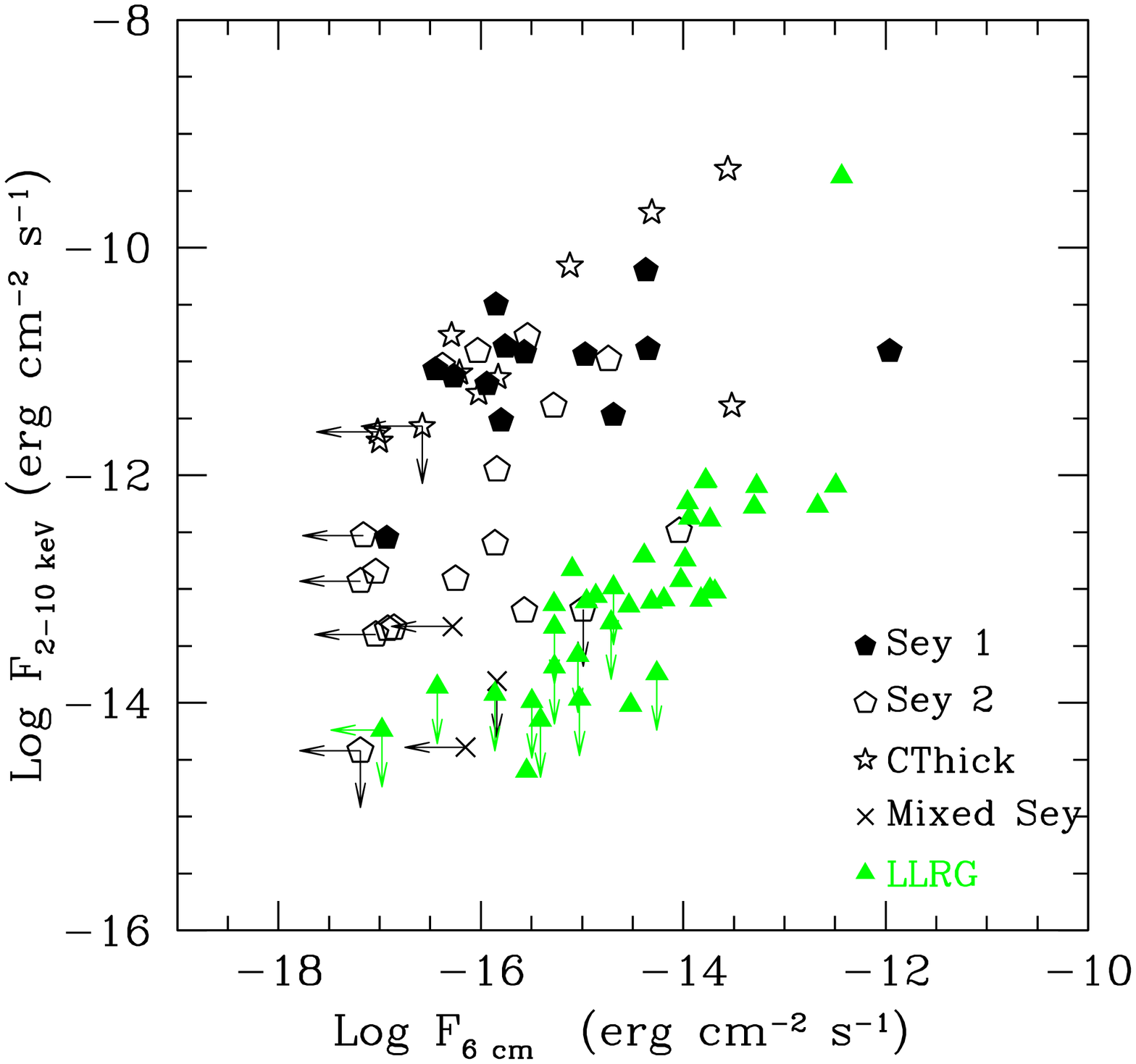}}
\caption{Left panel: Intrinsic 2-10 keV luminosity
versus core radio luminosity at 6 cm. Seyfert galaxies are plotted as polygons.  
Low Luminosity Radio Galaxies are plotted as solid triangles (Balmaverde \& Capetti 2006).
Seyferts and LLRGs best fit linear regression lines obtained using the Schmitt method have been plotted.
Right panel: Same plot of left panel in flux-flux space.}
\label{r2}
\end{center}
\end{figure*}   

In Figure~\ref{r2} (left panel) we plot
the nuclear 2-10 keV luminosity versus the core radio luminosity at 6~cm,
for the Palomar Seyfert sample (polygons) and the LLRGs (solid triangles).
The LLRGs best fit regression line has been obtained using the Schmitt method,
considering the censored data as reported in Table~\ref{stat}. Our best fit
is in agreement, within errors, with that reported in
Balmaverde \& Capetti (2006) obtained by excluding upper limits 
(with a slope of 1.02$\pm$0.10). Interestingly the group of Seyfert galaxies
with a radio excess emission discussed in the previous section, 
lay in the LLRGs correlation. 

The main result that comes from this analysis is that {\it the radio 
and X-ray luminosities in Seyfert galaxies
and LLRGs are significantly correlated, in addition both samples
show the same regression slope,
shifted by nearly three orders of magnitude}:

\begin{center}
- Seyferts: log L$_{X}$ = (0.97$\pm$0.01) log L$_{6 cm}$ + (5.23$\pm$0.28) 
 \end{center}

\begin{center}
 - LLRGs: log L$_{X}$ = (0.97$\pm$0.02) log L$_{6 cm}$ + (2.42$\pm$0.92)
\end{center}

The flux-flux plot in Figure~\ref{r2} (right panel),
shows a larger scatter, however the correlation remains statistically significant ($>$ 99.9\%)
meaning that the two correlations are not driven by distance effects,
as also demonstrated by the Kendall's $\tau$ test result (Table~\ref{stat}).

However, it must be taken into account that selection effects are probably 
introduced in the Seyfert and LLRGs samples. 
The Seyfert galaxy sample has been derived from the optical flux limited sample
of the Palomar survey (Ho, Filippenko, \& Sargent 1995). Originally,
the survey contained a sample of 60 Seyfert galaxies, which was indeed complete. 
We have X-ray data for 47 out of 60 sources,
thus we are probably introducing a slight bias toward high luminosity Seyfert nuclei.
The missing sources should have X-ray luminosities below 
log L$_{2-10 keV}$ $\sim$ 40.5 erg s$^{-1}$ (see Panessa et al. 2006).
A different selection procedure has been introduced when building
the core galaxy sample by Balmaverde \& Capetti (2006). These are early type
galaxies, initially selected from a flux limited VLA survey, which also have HST
and {\it Chandra} data available. This sample is not complete in any sense and is
probably biased towards high luminosity nuclei. Therefore, in both 
Seyferts and LLRGs samples, the bias at low luminosities could affect 
our results but only slightly, since the slopes of the correlations are calibrated
for over 8 orders of magnitude, while the lower luminosity sources should
account for only one or at most two orders of magnitude in the correlations.
In any case, this does not affect the observed dichotomy 
between Seyfert galaxies and LLRGs, i.e. putative
objects with an X-ray and a radio luminosity 
potentially populating the gap between the two correlations
are within the selection criteria of the two samples and therefore, if existing,
they should have been detected in both cases.
However, understanding the possible effects introduced 
by comparing a radio and an optical flux-limited sample is not a simple task.
Complete and bias-free samples are needed to overcome the present limitation.

In the literature, X-ray data, taken with ROSAT, have been correlated with 
6~cm radio core luminosities for a sample of low-power radio galaxies 
(Canosa et al. 1999). Interestingly, despite the use of 1 keV luminosity 
instead of the 2-10 keV one as in our case, the
best fit Schmitt slope of the regression line found by these authors is consistent 
within errors with our findings (slope = 1.15$^{+0.24}_{-0.23}$). Brinkmann et al. (2000)
correlated the ROSAT 2 keV monochromatic luminosity and the 6~cm luminosity for a sample of
radio-loud and radio-quiet galaxies, derived from a cross-correlation of 
the ROSAT All Sky Survey and the VLA 20 cm FIRST catalogue. 
The slope of the best fit regression line of radio-quiet
galaxies resulted to be steeper than those obtained for radio-loud objects (1.012$\pm$0.083 and
0.483$\pm$0.049, respectively). Note that
this sample is biased toward high luminosity sources, i.e. no radio-loud galaxies with 
log L$_{R}$ $\leq$ 40.5 erg s$^{-1}$ are present, and the regression line
is calibrated only over 4 orders of magnitude.
A soft X-ray versus radio (at 1.94 GHz) luminosity correlation has also been
observed in a sample of 93 AGN selected from the 
ROSAT Bright Survey (Salvato et al. 2004), with
a correlation slope which is in agreement with that found for 
XRBs, L$_{R}$ $\propto$ L$_{X}^{0.7}$ (Gallo et al. 2003). 

Here, the X-ray versus radio luminosity correlations found are valid
for nearly eight orders of magnitude in both Seyfert and LLRGs samples 
implying that (i) the X-ray and the radio source emission are strongly 
coupled down to very low luminosities and (ii) they follow a similar correlation slope
in the two samples suggesting either a common set of physical mechanisms producing
the emission or a combination of different mechanisms which end up
producing a similar correlation slope. 

There are severe limitations to our knowledge on the X-ray and radio source
emission both in radio-loud and radio-quiet AGN 
For example, the physical origin of X-ray emission in radio-loud AGN.
is a largely debated topic. On one hand, it has been proposed
that the nuclear X-ray emission is originated at the base of
a parsec-scale radio jet (e.g., Fabbiano et al. 1984). 
The flatness of the X-ray slope correlates with radio core dominance 
for radio-loud quasars, suggesting that a component of the X-ray emission is 
relativistically beamed (Zamorani et al. 1986, Shastri et al. 1993).
Moreover, their multi-wavelength emission is interpreted
in terms of synchrotron and inverse Compton jet models only 
(Chiaberge et al. 2003, Pellegrini et al. 2003).
Finally, the correlations between the radio core 
power and X-ray luminosities (Worrall 1997, Canosa et al. 1999, Brinkmann et al. 2001) 
observed for radio galaxies suggest that the X-ray spectrum is produced by
Compton up-scattering of the radio synchrotron photons from the relativistic jet. 
On the other hand, the detection of broadened FeK lines and short-term
variability (Gliozzi et al. 2004) support the idea that the emission
originates in an accretion flow. 
Evidence in favour of the presence of the two components
has been found in a few sources (Evans et al. 2004, Zezas et al. 2004, Sambruna et al. 2006).
 
Balmaverde \& Capetti (2006) have shown that LLRGs in their sample are
genuine active nuclei which host a radio-loud core. They found correlations 
between optical, X-rays and radio luminosities suggesting a common non-thermal
origin of the nuclear emission. They suggest that, given the low values of
the Eddington ratios obtained, the radiative output of these sources is 
most likely dominated by the jet, with only a small fraction due to 
the accretion process.

In the case of Seyfert galaxies, the most accepted physical scenario
for their X-ray spectra assumes that the emission is produced
by a disk-corona system, in which UV-soft photons 
from the accretion disk are comptonized and up-scattered into the hard X-ray band 
by a hot corona above the accretion disk (Haardt \& Maraschi 1991).
In agreement with this theoretical picture, X-ray studies of Seyfert galaxies have shown
that their intrinsic X-ray spectral and emission properties are
similar to those of more luminous AGN, including objects with very low luminosities down to 
L$_{X}$ $\sim$ 10$^{38}$ erg s$^{-1}$ (Ho et al. 2001, Terashima \& Wilson 2003, 
Cappi et al. 2006, Panessa et al. 2006).

\begin{figure*} 
\begin{center}
\parbox{18cm}{
\includegraphics[width=0.50\textwidth,height=0.35\textheight,angle=0]{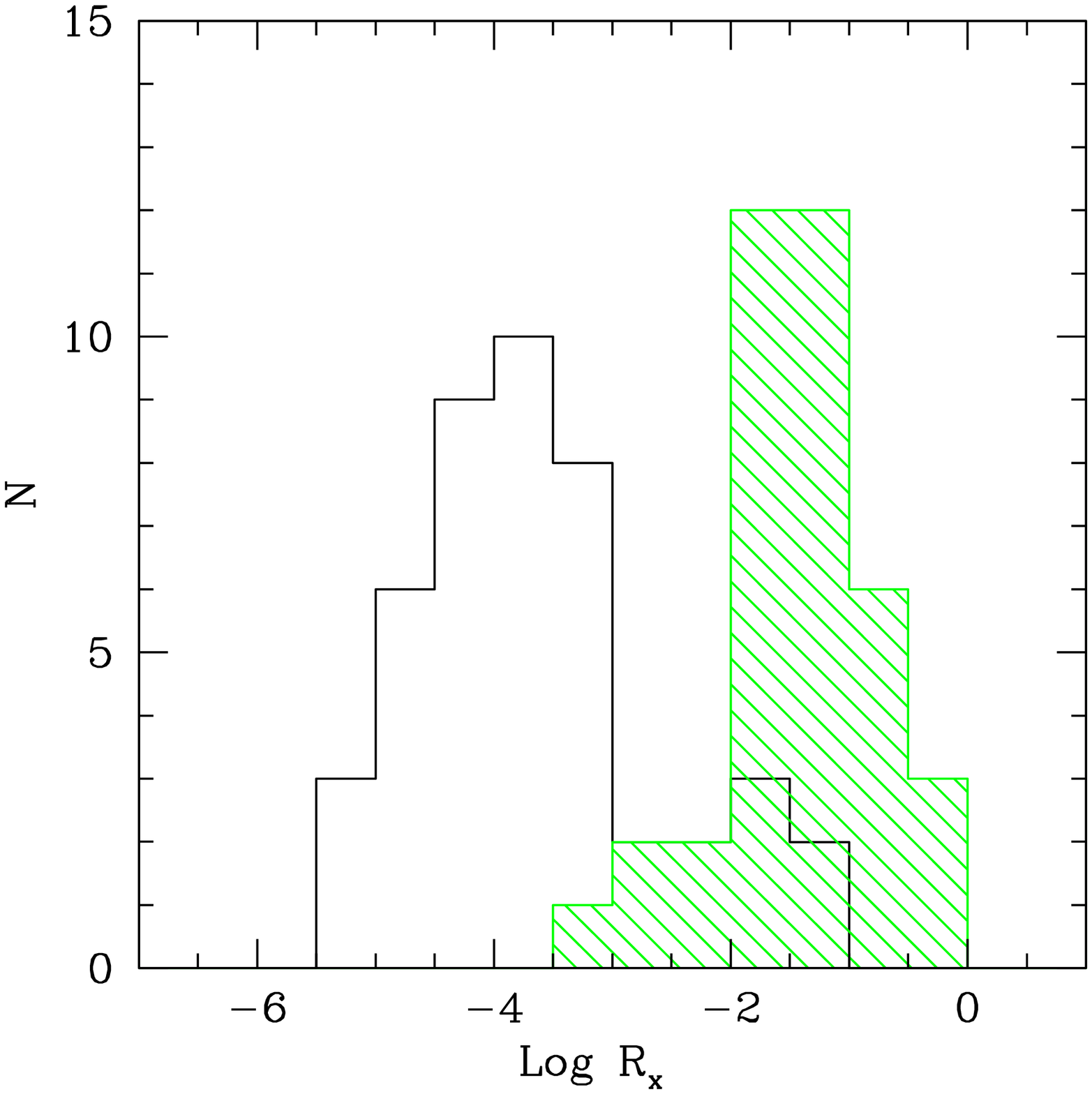}
\includegraphics[width=0.50\textwidth,height=0.35\textheight,angle=0]{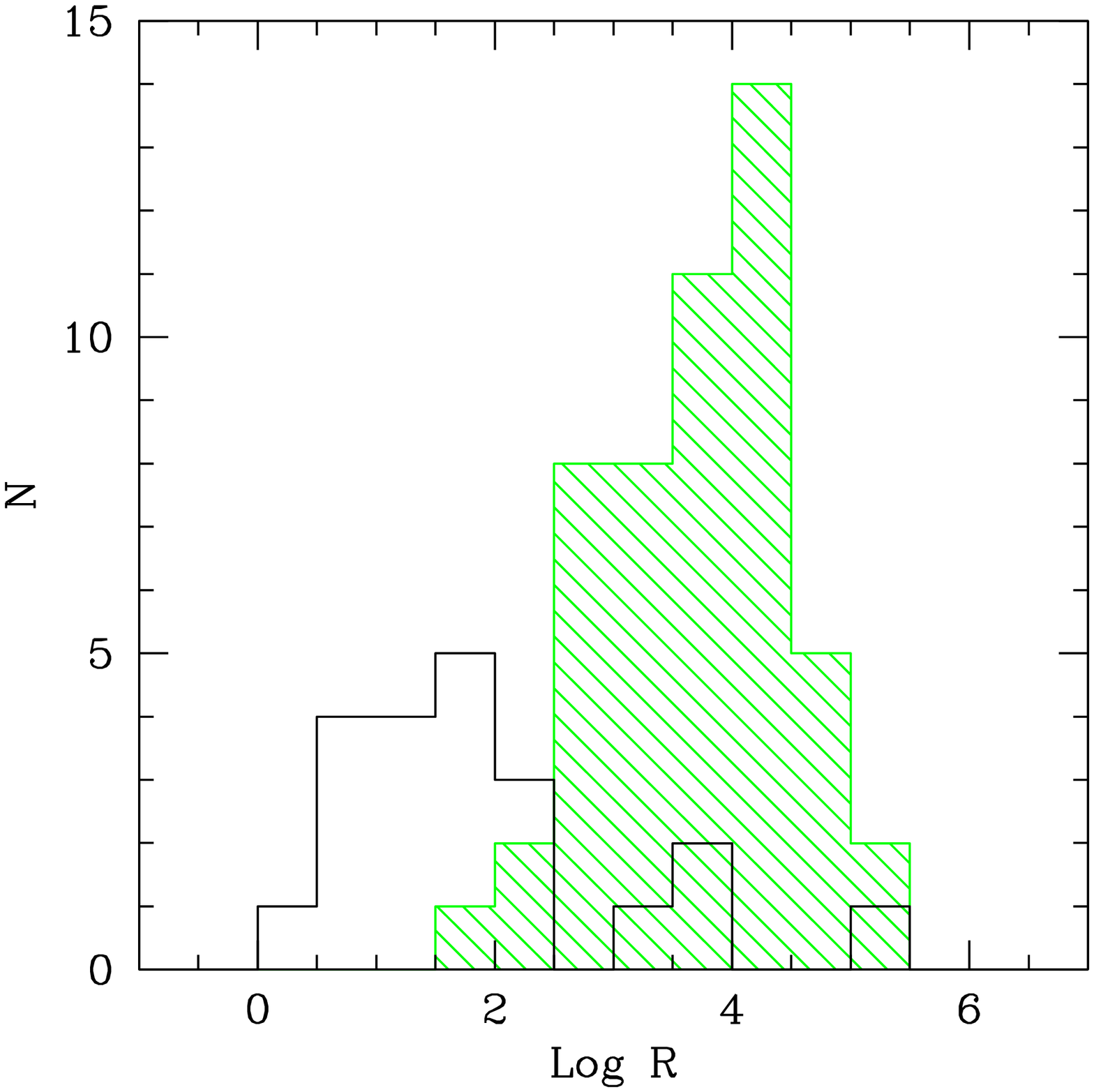}}
\caption{Left panel: Distributions of the R$_{X}$ $\equiv$ L$_{\nu}$(6~cm)/L(2-10 keV)
parameter for the Seyferts sample (left) and LLRGs sample (right, shaded histogram).
Right panel: Distributions of the R $\equiv$ L$_{\nu}$(6~cm)/L$_{\nu}$(B).}
\label{dist}
\end{center}
\end{figure*}   

For a given X-ray luminosity, 
the Palomar Seyfert galaxies are three orders of magnitude weaker in the radio band 
with respect to LLRGs. The origin of the radio emission 
and the physics of jets in Seyfert galaxies and LLAGN is not understood yet.
The radio parsec and sub-parsec scale study of a few LLAGN indicates that the 
radio emission is due to a synchrotron process from the base of a radio jet, as in
more luminous radio-loud AGN (Giroletti et al. 2005, Nagar et al. 2005).
Conversely, Bicknell (2002) suggests that there are major differences
between Seyfert and radio galaxies jets: Seyfert's jets are thermally 
dominated with sub-relativistic speeds while radio galaxies jets
are relativistic electron/positron flows. Another possible scenario has been proposed
by Ghisellini et al. (2004) to explain radio emission in Seyfert galaxies, 
such as the presence of 'aborted jets' responsible also for the emission in X-rays.
The latter theory could explain the
correlation found between the radio and X-ray luminosity in radio-quiet AGN. 

The correlations found here can be used,
when compared to theoretical predictions, to constrain the physical
parameters and improve our knowledge on how the disk-jet connection
works in AGN. 

\begin{figure*}[!htb] 
\begin{center}
\includegraphics[width=0.50\textwidth,height=0.35\textheight,angle=0]{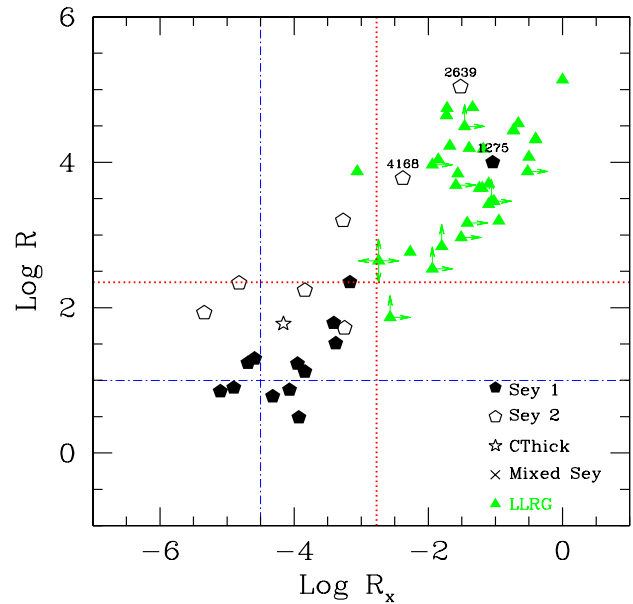}
\caption{Relation between the radio-loudness parameters 
R $\equiv$ L$_{\nu}$(6~cm)/L$_{\nu}$(B) and R$_{X}$ $\equiv$ L$_{\nu}$(6~cm)/L(2-10 keV)
for the Seyferts sample (polygons) and the LLRGs sample (triangles). Horizontal and vertical
lines represent the values where the maximum difference between the 
R$_{X}$ and R parameter distributions of Seyfert galaxies and LLRGs occurs.
Note that 3C028 has an upper limit at all frequencies measurements, therefore the radio-loudness
ratios cannot be determined.}
\label{rl}
\end{center}
\end{figure*}

\section{The radio-loudness}

Both the existence of a real dichotomy between radio-loud and 
radio-quiet AGN (not simply due to sample selection effects)
and its origin are still a matter of debate.
Seyfert galaxies are traditionally considered to be
radio-quiet objects, because of their low radio power
and the small radio to optical ratio (R $<$ 10). However, Ho \& Peng (2001)
have shown that, by considering their nuclear luminosities, most Seyfert galaxies 
are instead radio-loud AGN. In these sources,
the host galaxy emission plays an important role in the determination
of the photometric measurements, sometimes leading to a large overestimate
of the optical flux. Ho \& Peng (2001) measured the optical luminosities from HST images,
where the high spatial resolution allows to eliminate
the contribution from stellar light to the overall emission. They recomputed 
the radio and optical fluxes for a sub-sample of the Palomar Seyfert galaxies (mostly type 1-1.5)
and found that most of the objects analyzed are indeed radio-loud,
having their radio-loudness parameter R higher than the classical boundary 
value set at R=10 (Kellermann et al. 1994 and references therein). 
Note, however, that by instead using the luminosity criterion, 
P$_{6~cm}$ $\sim$ 10$^{25}$ W Hz$^{-1}$ sr$^{-1}$, only NGC~1275 results to be 
radio-loud in our sample. 

\begin{figure*} 
\begin{center}
\parbox{18cm}{
\includegraphics[width=0.50\textwidth,height=0.35\textheight,angle=0]{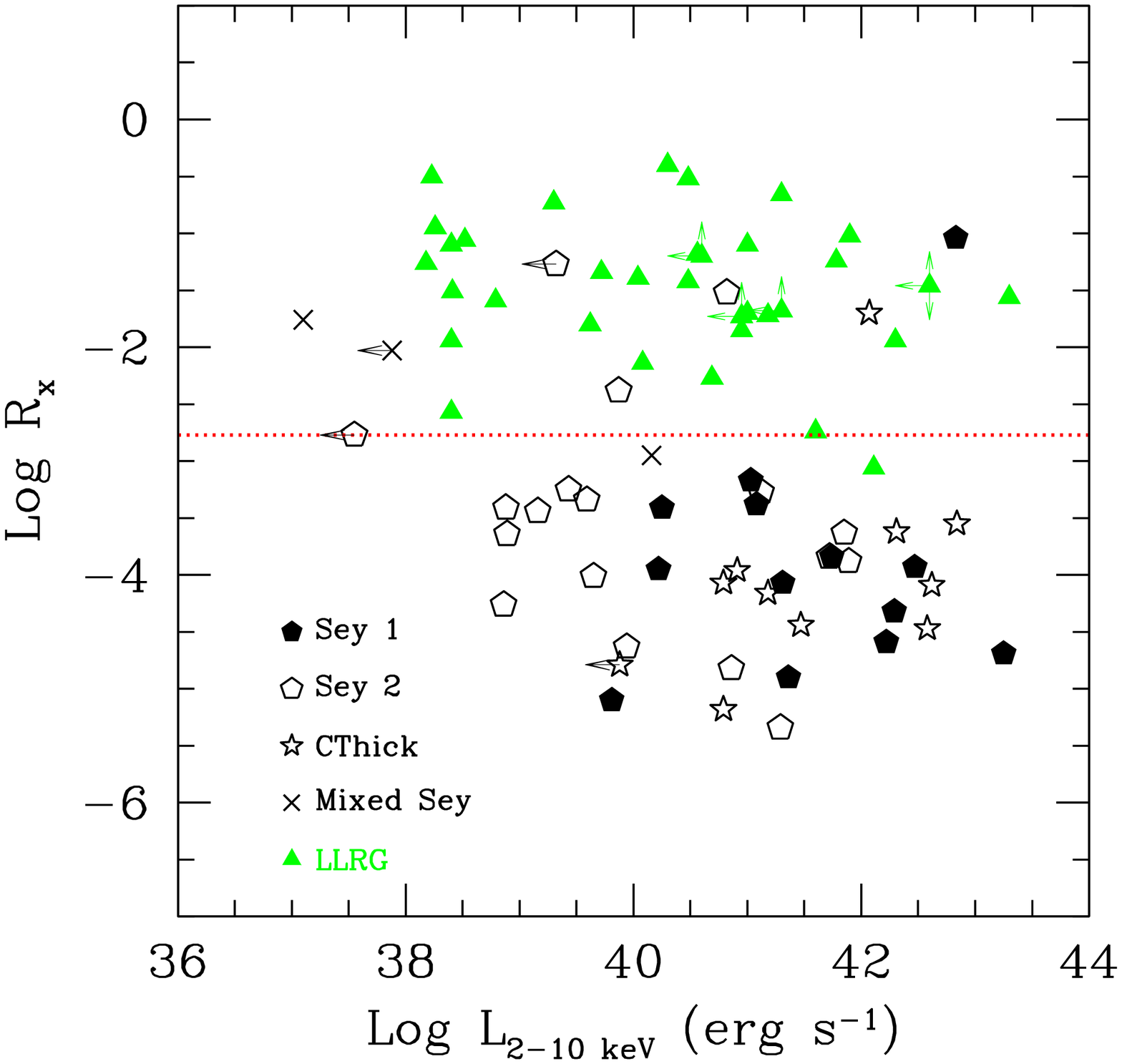}
\includegraphics[width=0.50\textwidth,height=0.35\textheight,angle=0]{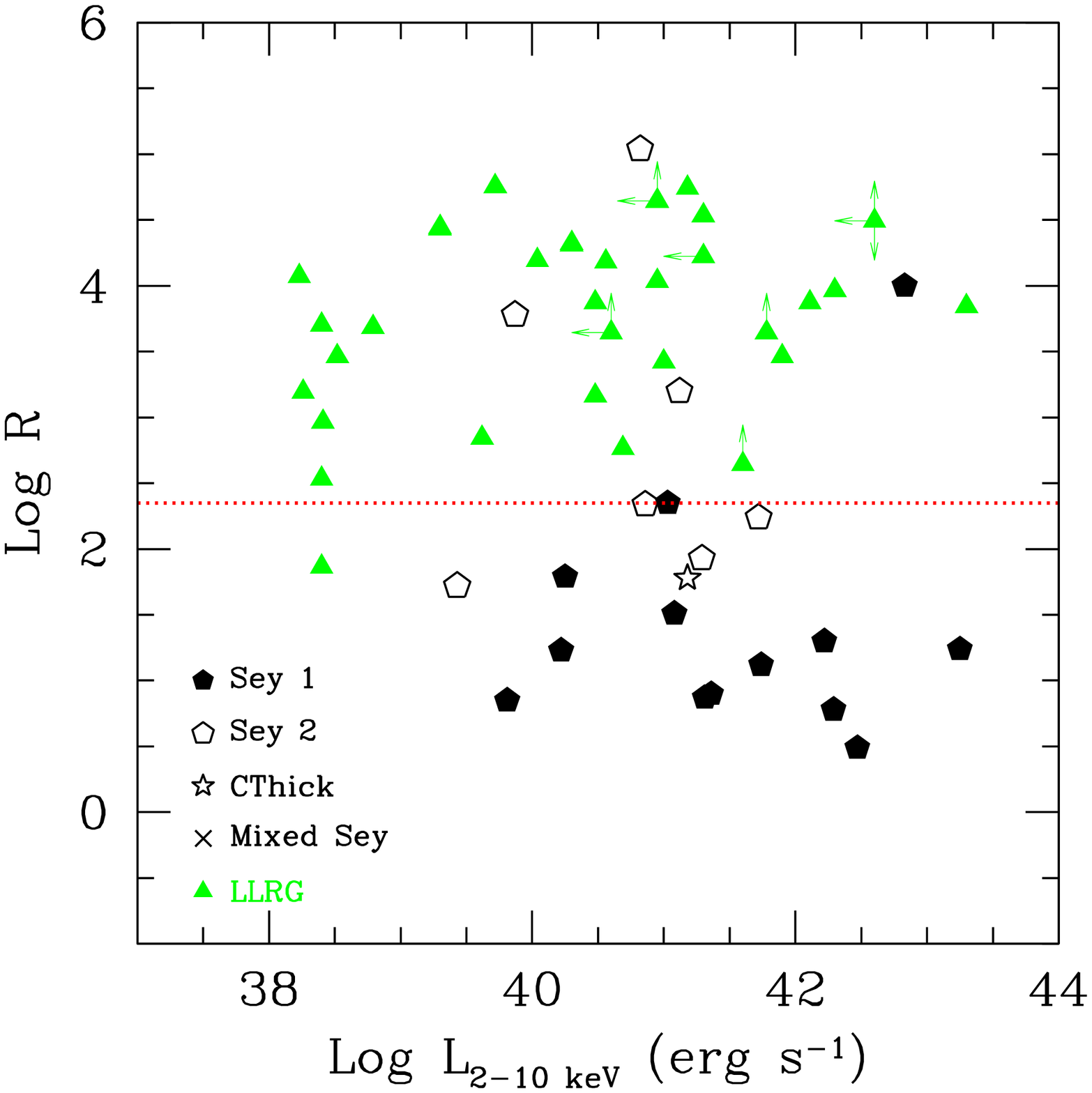}}
\caption{Left panel: R$_{X}$ $\equiv$ L$_{\nu}$(6~cm)/L(2-10 keV) 
and R $\equiv$ L$_{\nu}$(6~cm)/L$_{\nu}$(B) versus log L(2-10 keV) (erg s$^{-1}$), 
left and right panel respectively. The lines representing the values 
where the maximum difference between the are plotted respectively in the left and right panel.}
\label{rl_lum}
\end{center}
\end{figure*}   

We calculate the R parameter,
L$_{\nu}$(6~cm)/L$_{\nu}$(B), as shown in the distribution
of Figure~\ref{dist} (right panel). The optical data for LLRGs
have been taken from Balmaverde \& Capetti (2006), while R values for 
a sub-sample of type 1-1.9 Palomar Seyfert galaxies are taken from Ho \& Peng (2001). 
The mean value for LLRGs is 3.74$\pm$0.11 compared 
to 1.93$\pm$0.26 for the Seyfert galaxies and the two distributions
do not belong to the same parent population (KS probability
of 1.08$\times$10$^{-8}$).

Recently, Terashima \& Wilson (2003) have introduced a new definition of 
the radio loudness parameter by comparing the 6~cm radio luminosity 
to the 2-10 keV luminosity, R$_{X}$ $\equiv$ L$_{\nu}$(6~cm)/L(2-10 keV).
The use of the X-ray luminosity with respect to the optical one
should largely avoid extinction problems which normally occur in the optical band
and which could cause an overestimation of R.
We therefore calculated the R$_{X}$ parameter for both the Seyfert galaxies and the 
LLRGs sample (see the distributions of the two samples in
Figure~\ref{dist}, left panel). A Kolomogorov-Smirnov test,
computed by excluding upper limits,
results in a probability of 3.5$\times$10$^{-13}$
that the two samples are drawn from the same parent population.
The mean log R$_{X}$ value of LLRGs is 
-1.40$\pm$0.11 compared to -3.64$\pm$0.16 of Seyfert galaxies.

The two classes of sources show clearly different distributions
in both the R$_{X}$ and the R parameters, as also clear in Figure~\ref{rl}
where we plot the log R versus log R$_{X}$. 
We have compared the Seyferts and LLRG cumulative distributions for the R$_{X}$ parameter
by using the KS test and found that the maximum separation between the two distributions
(D = 0.81) is located at R$_{X}$=-2.755$\pm$0.015.
In the case of the R parameter, the maximum distance (D = 0.77) 
is found at R=-2.400$\pm$0.050.
We plotted the two boundaries as vertical and horizontal dotted lines in Figure~\ref{rl}.

\begin{figure*} 
\begin{center}
\includegraphics[width=0.50\textwidth,height=0.35\textheight,angle=0]{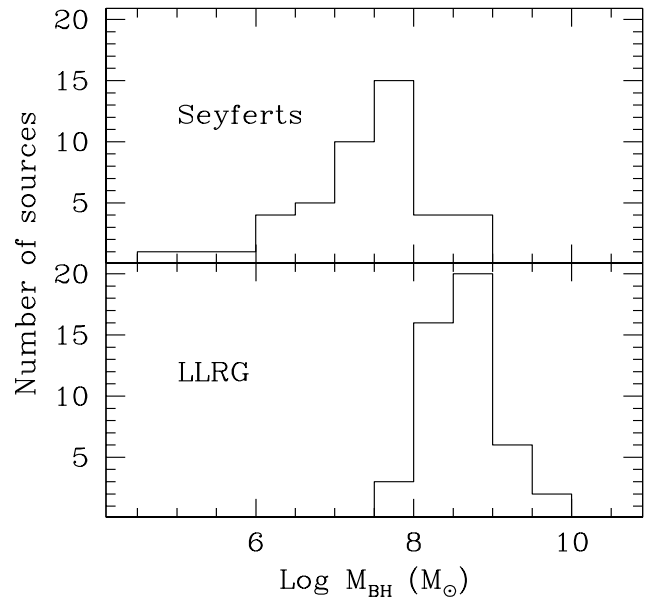}
\caption{Black hole mass distribution for Seyfert galaxies (top)
and LLRGs (bottom).}
\label{mbh}
\end{center}
\end{figure*}   

Terashima \& Wilson (2003) have compared the R and the R$_{X}$ 
radio-loudness parameters for a sample of LLAGN, Seyfert galaxies
and PG quasars, deriving a relation between the two variables 
(log R = 0.88 log R$_{X}$ + 5.0). 
We also calculated the best fit linear regression line: 
log R = 0.91$\pm$0.06 log R$_{X}$ +5.28$\pm$0.18.
Within errors, the latter is consistent with that found by Terashima \& Wilson (2003).
These authors used that relationship to derive the boundary between 
the radio-loud and radio-quiet objects in R$_{X}$, i.e. by fixing R=10, they
obtained log R$_{X}$ $=$ -4.5. According to the above boundaries, most Seyfert galaxies
should be considered as radio-loud AGN, as well as all the LLRGs. 
However, if we take into account
the radio-loud nature of LLRGs (Balmaverde \& Capetti 2006) and consider 
the different radio powers of Seyfert galaxies with respect to LLRGs, we could redefine the boundaries between 
radio-loud and radio-quiet AGN as above: R$_{X}$=-2.755$\pm$0.015 and R=-2.400$\pm$0.050.
The validity of such limits should be tested in the future with larger samples 
of radio-quiet and radio-loud AGN. These different boundaries found for low luminosity
active nuclei with respect to those of luminous AGN could suggest a dependence
of R and R$_{X}$ with luminosity. However, as shown in Figure~\ref{rl_lum} 
where we plot the radio loudness parameters,
R$_{X}$ (left panel) and R (right panel), versus the X-ray luminosity, there is
no evidence of such trend within the low luminosities spanned by our sample.

\section{Radio-loudness versus M$_{BH}$ and Eddington ratio}

Several recent works have attempted
to interpret the occurrence of radio-loud nuclei in the center
of only few percent of AGN as related to the host galaxy properties.
For example, it has been observed that spiral galaxies preferentially 
harbour radio-quiet AGN, while early-type galaxies host both
radio-quiet and radio-loud AGN (Hutchings 1983). Actually, LLRGs
have been selected to be early-type galaxies while only few Seyfert galaxies
are hosted by elliptical host galaxies (e.g., NGC~3608, NGC~4168, NGC~4472 and NGC~6482).
Moreover, Dunlop et al. (2003) have found that radio-loud AGN are associated with the 
most massive black hole hosts, providing a natural explanation for why radio-quiet outnumber
radio-loud AGN and suggesting that the radio output is a strong function of the
black hole mass.

\begin{figure*} 
\begin{center}
\parbox{18cm}{
\includegraphics[width=0.50\textwidth,height=0.35\textheight,angle=0]{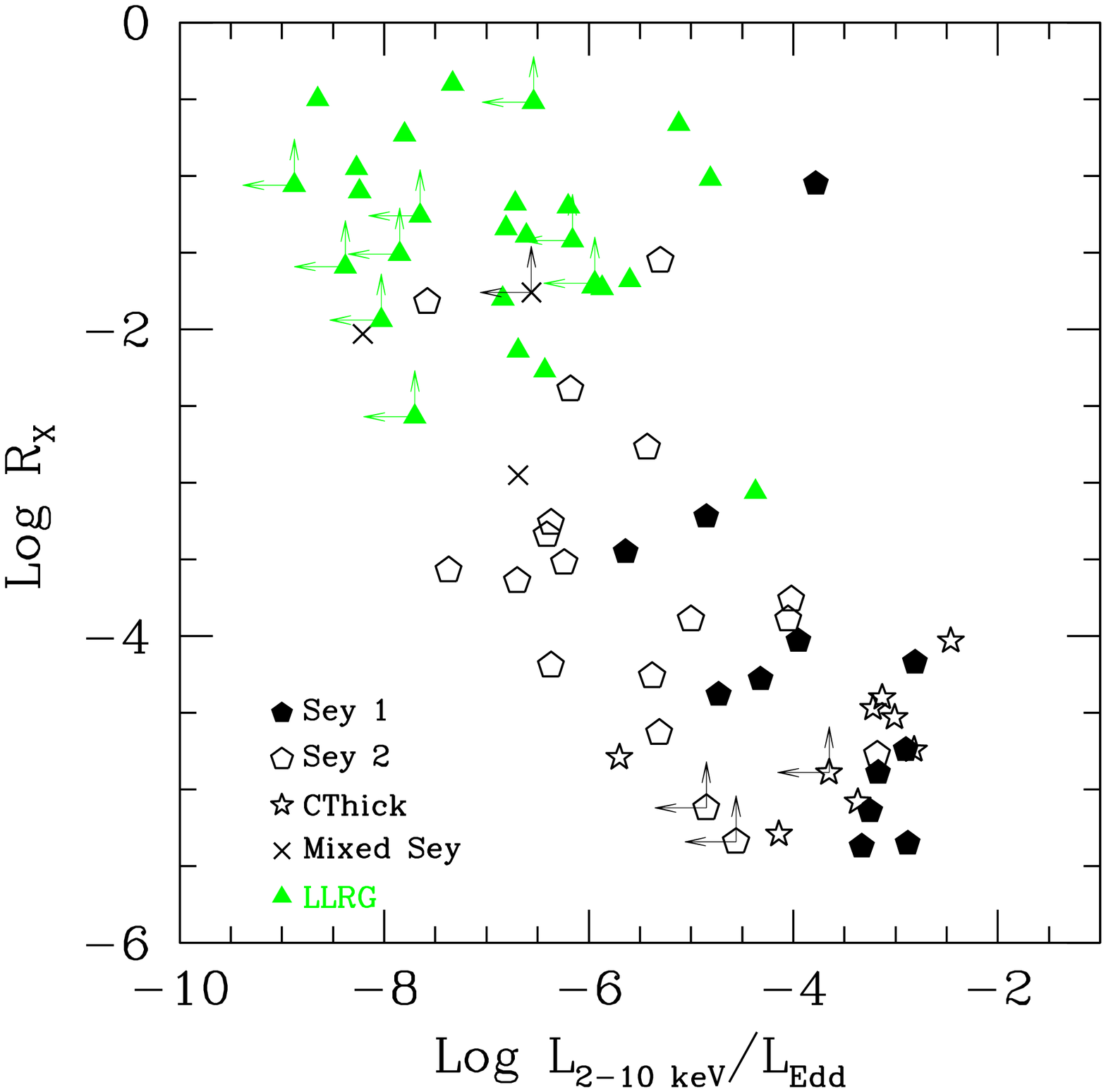}
\includegraphics[width=0.50\textwidth,height=0.35\textheight,angle=0]{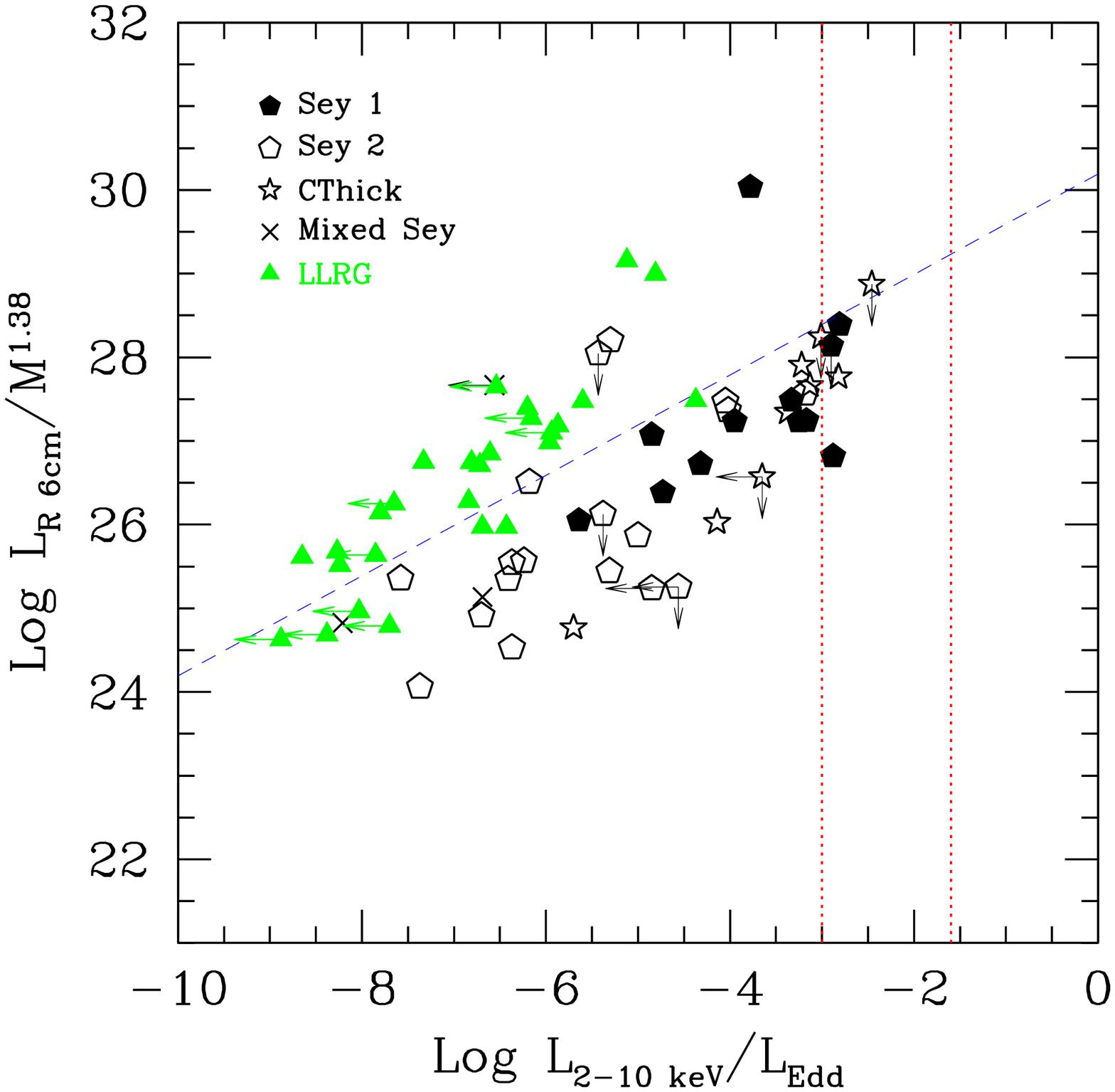}}
\caption{Left panel: X-ray radio loudness versus Eddington ratio 
L$_{2-10 keV}$/L$_{Edd}$. Right panel: Radio luminosity and M$_{BH}^{1.38}$ ratio
(from the fundamental plane equation in Merloni, Heinz \& Di Matteo 2003) versus the
Eddington ratio.}
\label{edd}
\end{center}
\end{figure*}   

In Figure~\ref{mbh} we present the black hole mass distribution 
for the sample of Seyfert galaxies (top) and LLRGs (bottom).
The M$_{BH}$ estimates for the Seyfert galaxies
have been taken from the literature, where they were obtained using
a variety of methods (Panessa et al. 2006).
Palomar Seyfert M$_{BH}$ are distributed
from $\sim$ 10$^{5}$ to 10$^{8}$ M$_{\odot}$
with a peak at 10$^{7-8}$ M$_{\odot}$.
The M$_{BH}$ distribution of LLRGs has already been shown in
Capetti \& Balmaverde (2006): core galaxies and low luminosity
radio galaxies in their sample show a similar M$_{BH}$ distribution
and a narrow range of values, i.e. log M$_{BH}$ $\sim$ 7.5-9.5.
Compared to our Seyfert galaxy sample, there is an evident lack of objects
with log M$_{BH}$ $<$ 7.5. Note that most
of the M$_{BH}$ estimates have been obtained
by using the mass vs. stellar velocity dispersion relation.
Woo \& Urry (2002) have compared M$_{BH}$ estimates derived
from direct and indirect measurements of the
stellar velocity dispersions for a sample of radio galaxies.
The two methods agree on average to within 10\%, however the rms scatter is a factor of 4.
In this work, radio galaxies are mainly hosted by massive black hole hosts,
even though the Seyfert and LLRGs distributions overlap. 
This is in agreement with the results found by Sikora et al. (2006),
Laor (2003) and Wu \& Han (2001). However, the role played by selection effects 
on this is still unclear. Indeed large compilations of data have shown that
M$_{BH}$ for radio-quiet and radio-loud objects
show similar distributions, with several radio-loud objects having
M$_{BH}$ $\leq$ 10$^{8}$ M$_{\odot}$ (Woo \& Urry 2002, Ho 2002, Wang et al. 2006).

The relation between radio loudness and host properties
has been pointed out also by Capetti \& Balmaverde (2006). They
have found that the radio-loudness parameter
in AGN hosted by early-type galaxies depends on the host's brightness profile,
where core galaxies only host radio-loud AGN and radio-quiet AGN are found in 
power-law galaxies. Since the brightness profile is determined
by the galaxy's evolution and merger history, the radio-loudness
is therefore related to those basic parameters such as mass and spin 
which are probably triggered by mergers. 

An important role is played also by the accretion rate (Laor 2003). 
Originally, Ho (2002) found a trend of increasing 
radio loudness with decreasing Eddington ratio.
In the left panel of Figure~\ref{edd} we show the radio loudness
versus the Eddington ratio expressed as L$_{2-10 keV}$/L$_{Edd}$. The Seyfert and LLRGs
seem to confirm this trend which could suggest that the formation of a jet in LLAGN
is related to the accretion rate similarly to what is observed in XRBs.
However Sikora et al. (2006) by using samples of both high and low luminosity AGN,
found two separate trends followed by
radio-quiet and radio-loud AGN, and proposed a model which includes the black hole spin
as one of the fundamental parameters, together with the mass, that determines
the radio loudness in AGN, as also previously conjectured 
by other authors (Blandford 1999, Meier 1999).

In the XRBs low/hard state the intermittency
of the jet production is connected to the accretion rate (Fender, Belloni \& Gallo 2004).
Apparently, the similarity between the jet and flow energy production in XRBs and AGN is
rejected by the slope of the correlation we found for Seyfert galaxies and LLRGs
which differs from those found by Gallo, Fender \& Pooley (2003)
for low/hard state XRBs, L$_{R}$ $\propto$ L$_{X}^{0.7}$.
However, several authors were motivated to unify the central engine
mechanisms in Galactic Black Holes (GBHs) and super massive accreting black holes (SMBHs) 
(Maccarone, Gallo \& Fender 2003, Merloni, Heinz \& Di Matteo 2003, Falcke et al. 2004).
The observation of a correlation between the X-ray, radio and black hole
mass valid both for GBH and SMBH (the so-called 'fundamental plane of black
hole activity') has different interpretations. Merloni, Heinz \& Di Matteo (2003) 
suggest that the X-ray emission is produced by a radiatively inefficient
accretion flow (ADAF) while the radio emission originates in a relativistic jet.
Given the observed correlations, the flow and the jet are strongly coupled.
The fundamental plane equation also implies that the radio loudness should
increase as the accretion rate decrease and that the radio-loud
and radio-quiet dichotomy should appear only at high accretion rates,
similarly to what is observed in the high/very high transition in GBH.

In Figure~\ref{edd} (right panel), we plot the ratio between
the radio luminosity and M$_{BH}^{1.38}$ (the latter scaling
is derived from the fundamental plane equation in Merloni, Heinz \& Di Matteo 2003) versus
the Eddington ratio.
The objects considered in the Merloni, Heinz \& Di Matteo (2003) sample,
collapse into a single track in this space 
(see Figure~7 in Merloni, Heinz \& Di Matteo 2003). In our case, LLRG and
the Seyfert galaxy follow two separate tracks indicating that a dichotomy 
is still observed down to low Eddington ratios.
The region between the two vertical lines correspond to the L$_{2-10 keV}$/L$_{Edd}$
values where the switch in the accretion mode is expected and where the radiatively
efficient regime is dominant. According to this model, most of the
Seyfert galaxies and all LLRGs should accrete in an inefficient radiative regime.
Despite that few Seyferts populate the 'efficient accretion' zone,
we do not observe any hint of a switch in the accretion mode above
L$_{2-10 keV}$/L$_{Edd}$ $>$ 10$^{-3}$. Moreover most Seyfert galaxies of this sample, 
in particular those of type 1, have spectral
characteristics inconsistent with models of low
radiative efficiency, such as an X-ray power-law slope steeper than those predicted
by ADAF models and/or the presence of an FeK line which accretion thin flows are unable to produce.

We over-laid the fundamental plane equation by Merloni, Heinz \& Di Matteo (2003) 
(short-dashed line in Figure~\ref{edd}, right panel) derived as a function of log L$_{2-10 keV}$/L$_{Edd}$. 
It appears to be a good fit for the whole the Seyfert galaxy plus LLRG sample. 
However, Seyfert galaxies and LLRGs seem to show two different fundamental planes. 
Indeed, Wang et al. (2006) have come to the same conclusion by investigating
a sample of broad line SDSS AGN, both radio-quiet and radio-loud.
They have compared fluxes at 1.4 GHz and 0.1-2.4 keV with 
M$_{BH}$ measurements. On the basis of their results these authors
have suggested that radio-loud sources
may obey the same correlation as radio-quiet sources, only
shifted by a different factor depending on the radio-loudness parameter.
They also suggest that the beaming effects alter significantly
the radio luminosity in radio-loud AGN, concluding that the fundamental plane
in these sources is apparently unreliable if this effect cannot be removed. 

\section{Summary}

We have investigated the correlation bewteen X-ray and radio luminosities for a
sample of Seyfert galaxies and low luminosity radio galaxies (LLRGs)
having nuclear data both at X-ray and radio frequencies. The radio loudness
parameter and its dependence on luminosity, black hole mass and Eddington ratio
have also been discussed. We summarize our results as follows:

\begin{itemize}

\item Nuclear 2-10 keV X-ray and core radio luminosities at 20~cm, 6~cm and 2~cm 
are significantly correlated in the sample of nearby Seyfert galaxies 
down to very low luminosities
(L$_{2-10 keV}$ $>$ 10$^{38}$ erg s$^{-1}$), suggesting a strong
coupling between the X-ray and the radio emission mechanisms. The former is
commonly ascribed to a disk-corona system and the latter is 
likely associated with a core jet/outflow. 

\item The X-ray versus 6~cm radio luminosity correlations hold
for nearly eight orders of magnitude in both Seyfert and LLRGs samples. 
Interestingly in the two samples we found a similar correlation slope 
(L$_{X}$ $\propto$ L$_{R}^{0.97}$)
suggesting either common physical mechanisms in their nuclei 
or a combination of different mechanisms which end up
producing a similar spectral slope. It has been suggested that 
the X-ray emission in LLRGs
could have a common non-thermal origin as the radio emission, as for example due to 
synchrotron radiation from a relativistic jet (Balmaverde et al. 2006);
however it is unlikely
that this is also valid for Seyfert galaxies, for which a contribution 
from the accretion-flow/hot-corona system is expected in X-rays.

\item Both the classical radio loudness parameter and the X-ray one 
show a distribution in Seyfert galaxies which is different
from that of LLRGs. According to the classical boundaries between 
radio-loud and radio-quiet AGN, most of the Palomar Seyfert galaxies
and all LLRGs are on the radio-loud side. However, if one redefines the
boundaries in both parameters (R$_{X}$=-2.755$\pm$0.015 and R=-2.400$\pm$0.050) 
the difference in the radio power between the two classes of objects is recovered. 
The boundaries are then different with respect to those of luminous AGN
and this could be ascribed to 
an intrinsic property of low luminosity AGN. We find however no dependence
of the radio loudness parameter on the luminosity.

\item Black hole masses in the Seyferts and LLRGs samples
are differently distributed. Early-type LLRGs host more massive
black holes than Seyfert galaxies, which commonly reside in spiral
galaxies. This result confirms previous findings (Laor 2000, Wu \& Han 2001, Sikora et al.
2006) that higher black hole masses are associated to higher radio loudness parameters.
However, the radio-loudness is not only 
host galaxy morphology dependent, but other parameters 
should play a role such as the host's brightness profile (Capetti \& Balmaverde 2006),
the accretion rate and the black hole spin (Blandford 1999, Meier 1999, Sikora et al. 2006). 

\item We confirm an anti-correlation between the 
radio loudness and the Eddington ratio, i.e. AGN with lower
L$_{2-10 keV}$/L$_{Edd}$ tend to be more radio-loud. 
There is however no clear evidence of a transition between a radiatively
efficient to a radiatively inefficient accretion regime and 
the radio-loud/radio-quiet dichotomy seems to hold down to low 
luminosities and low Eddington ratios, implying that the accretion
rate is not the unique parameter which triggers the radio loudness.

\end{itemize}

\begin{acknowledgements}
We thank Mauro Dadina for helpful suggestions.
F.P. acknowledges support by a "Juan de la Cierva" fellowship.
Financial support for F.P., X.B. and F.J.C. was provided by the Spanish
Ministry of Education and Science, under project ESP2006-13608.
\end{acknowledgements}

\end{document}